\documentclass[%
 reprint,
 superscriptaddress,
 showpacs,
nobibnotes,
 amsmath,amssymb,
 aps,
 prb,
floatfix,
]{revtex4-1}

\usepackage{graphicx}
\usepackage{hyperref}

\newcommand{\vect}[1]{\vec{#1}}
\newcommand{\kB}{\ensuremath{\text{k}_\text{B}}}
\newcommand{\dd}[2]{\ensuremath{\frac{\text{d}#1}{\text{d}#2}}}

\newcommand{\refeq}[1]{Eq.~(\ref{#1})}
\newcommand{\reffig}[1]{Fig.~\ref{#1}}
\newcommand{\refsfig}[2]{\reffig{#1}(#2)}

\begin{document}

\preprint{APS/123-QED}

\title{Twist in the bias-dependence of spin-torques in magnetic tunnel junctions}

\author{S\"oren Boyn}
\author{Jo\~ao Sampaio}%
\altaffiliation[Present address: ]{Laboratoire de Physique des Solides, CNRS, Univ.\ Paris-Sud, Universit\'e Paris-Saclay, 91405 Orsay, France.}
\author{Vincent Cros}%
\author{Julie Grollier}\email{julie.grollier@thalesgroup.com}%
\affiliation{Unit\'e Mixte de Physique, CNRS, Thales, Univ.\ Paris-Sud, Universit\'e Paris-Saclay, 91767 Palaiseau, France.}%

\author{Akio Fukushima}
\author{Hitoshi Kubota}
\author{Kay Yakushiji}
\author{Shinji Yuasa}
\affiliation{National Institute of Advanced Industrial Science and Technology (AIST), Spintronics Research Center, Tsukuba, Ibaraki 305-8568, Japan.}%

\date{\today}

\begin{abstract}
The spin-torque in magnetic tunnel junctions possesses two components that both depend on the applied voltage. Here, we develop a new method for the accurate extraction of this bias-dependence from experiments over large voltage ranges. We study several junctions with different magnetic layer structures of the top electrode. Our results obtained on junctions with symmetric CoFeB electrodes agree well with theoretical calculations. The bias-dependences of asymmetric samples, with top electrodes containing NiFe, however, are twisted compared to the quadratic form generally assumed. Our measurements reveal the complexity of spin-torque mechanisms at large bias.
\end{abstract}

\maketitle


\section{Introduction}

The effect of spin-torque\cite{slonczewski_1996_current-driven,*berger_1996_emission,*slonczewski_2005_currents,ralph_2008_spin} is widely used to control the magnetization of the free layer in spin valves and magnetic tunnel junctions.\cite{myers_1999_current-induced,*huai_2004_observation,*kubota_2005_evaluation,stiles_2006_spin-transfer} It can induce both stable switching and dynamic oscillations.\cite{kiselev_2003_microwave} Owing to their high tunneling magnetoresistance (TMR),\citep{parkin_2004_giant,*yuasa_2004_giant} MgO-based magnetic tunnel junctions are promising candidates for applications to memory devices,\cite{chappert_2007_emergence,*diao_2007_spin-transfer,*parkin_2008_magnetic,*parkin_2015_memory,*kent_2015_new} oscillators,\cite{dussaux_2010_large,*deac_2008_bias-driven} logic,\cite{kruglyak_2010_magnonics} and neural networks.\citep{locatelli_2014_spin-torque} The mechanism underlying spin-torque is an essential part of these techniques. However, the dependence of spin-torque on the applied bias shows a wide influence of material parameters that have not allowed to draw a full picture of this effect yet.\cite{brataas_2012_current-induced}

In general, spin-torque in magnetic tunnel junctions has two components: one acting as a damping or anti-damping depending on the direction of the current, the other one acting similarly to a magnetic field along the polarizer magnetization axis.\citep{theodonis_2006_anomalous} The damping-like torque (DL) has first been predicted and observed in spin valve structures where it increases linearly with current.\citep{slonczewski_1996_current-driven,katine_2000_current-driven,*grollier_2001_spin-polarized} In tunnel junctions, however, the presence of the tunneling barrier leads to a strong filtering of the electron wave vectors so that only electrons from a small fraction of the Fermi surface contribute to the tunneling current.\citep{butler_2001_spin-dependent} As a result, the electron spin dephasing length is strongly increased. This gives rise to the field-like torque (FL). Measurements have revealed that its amplitude can be of the same order as that of the damping-like torque.\citep{sankey_2007_measurement,*kubota_2007_quantitative,petit_2007_spin-torque,ralph_2008_spin}

For the case of symmetric magnetic tunnel junctions with identical top and bottom electrodes, theoretical calculations based on different methods unanimously yield a quasi-linear bias-dependence of the damping-like torque and a quadratic bias-dependence of the field-like  torque.\citep{theodonis_2006_anomalous,xiao_2008_spin-transfer,manchon_2008_description,wilczynski_2008_free-electron,chshiev_2008_voltage,heiliger_2008_ab,kalitsov_2009_spin-transfer,franz_2013_influence} These dependences have also been measured experimentally.\citep{kubota_2007_quantitative,sankey_2007_measurement,wang_2009_bias,jung_2010_bias} For the case of asymmetric magnetic tunnel junctions with different compositions of the electrodes, however, theoretical results that take into account the different electronic band structures predict a large variety of bias-dependences.\citep{tang_2009_controlling,tang_2010_influence,manchon_2010_signatures,datta_2012_voltage,kalitsov_2013_spin} In general, the damping-like torque is assumed to gain a quadratic term whereas the field-like torque becomes linear at small bias. However, there are few experimental measurements\citep{chanthbouala_2011_vertical-current-induced,oh_2009_bias-voltage} testing these predictions in junctions with different electrodes, and many questions remain unanswered. In particular the bias-dependence of spin-torques in junctions with composite magnetic layers has never been studied.

In order to probe the bias-dependence experimentally several techniques have been developed.\cite{sankey_2007_measurement,*kubota_2007_quantitative,oh_2009_bias-voltage,park_2011_measurement,xue_2012_network,wang_2011_time-resolved} However, they are either limited to low voltages and demand high-frequency measurements or rely on critical external parameters. Here, we present a new technique to determine the bias-dependences of spin-torques. It is based on simple dc measurements as a function of applied magnetic field and bias. Thanks to the TMR effect, the relative magnetization of the two electrodes is derived from the resistance. This allows to create the phase diagram of each junction from which we determine the spin-torque evolutions with voltage.\citep{oh_2009_bias-voltage,park_2011_switching} For some special cases of bias-dependence, the critical fields and voltages for magnetization reversal can be calculated analytically.\citep{li_2008_perpendicular,oh_2009_bias-voltage,bernert_2014_phase} However, no analytical solution exists in the general case. We therefore calculate the phase diagram numerically, which leaves us free from any restrictions on the bias-dependence of spin-torques and enables us to reveal dependencies beyond usual assumptions. The comparison of these numerical simulations to the experimental phase diagrams finally allows for the determination of the bias-dependent spin-torques.

We studied tunnel junctions with different top electrode compositions while keeping the same bottom electrode. The samples cover the range from a fully symmetric structure with electrodes of the same materials, over a sample with identical interfaces but different materials in the bulk, to entirely different electrode materials. The results of the bias-dependent torques in these samples reveals the need for a more complex description of spin-torques than commonly assumed.

\section{Simulations}

The simulations of the phase diagrams are based on a macro-spin model of the normalized free-layer magnetization $\vect{m} = \vect{M} / M_\mathrm{s}$. We describe its time-dependent dynamics using the Landau--Lifshitz--Gilbert (LLG) equation with spin-torques:\citep{slonczewski_1996_current-driven,zhang_2002_mechanisms}
\begin{multline}
	\dd{\vect{m}}{t} \, = \, - \gamma_0 \, \vect{m} \times \mu_0 \vect{H}_\mathrm{eff} + \alpha \, \vect{m} \times \dd{\vect{m}}{t}\\
	+ \gamma_0 \, T_\mathrm{IP} \, \vect{m} \times (\vect{m} \times \vect{p}) - \gamma_0 \, T_\mathrm{OOP} \, \vect{m} \times \vect{p} \text{ .} \label{eqn_llg_stt}
\end{multline}
Here, the effective magnetic field $\vect{H}_\mathrm{eff}$ includes the externally applied field $\vect{H}_\mathrm{ext}$ as well as the contribution from the shape anisotropy of our films: approximating the thin free layer as an ellipsoid, the latter is taken into account by a damping-like uniaxial anisotropy field $\vect{H}_\mathrm{an}$ and a demagnetizing field $H_\mathrm{d} = \mu_0M_\mathrm{s} N_\mathrm{z}$ with demagnetization factor $N_\mathrm{z}$ perpendicular to the free layer.\citep{osborn_1945_demagnetizing} We simulate the influece of finite temperature by a three-dimensional thermal fluctuation field $\mu_0 \vect{H}_\mathrm{T}$ at each integration step. This field's Cartesian component amplitudes are chosen independently from a Gaussian distribution of variance $\sigma_\mathrm{T} = \sqrt{2 \alpha \kB T / (\gamma_0 M_\mathrm{s} V \Delta t)}$ with Gilbert damping constant $\alpha$, Boltzmann constant $\kB$, temperature $T$, absolute electron gyromagnetic ratio $\gamma_0$, saturation magnetization $M_\mathrm{s}$, volume $V$, and time step $\Delta t$.\citep{brown_1963_thermal} The fixed-layer magnetization $\vect{p}$ that determines the polarization of the tunneling electrons is kept constant along the easy axis of the free layer. Its amplitude is set to $|\vect{p}| = 1$, normalizing the resulting torque amplitudes.

The amplitudes of damping-like and field-like torque are functions of applied bias $V$ that are expressed as Taylor polynomials:
\begin{subequations}
	\label{eqn_stt_bias}
	\begin{alignat}{3}
		&T_\mathrm{IP}{}(V) \, &= \, a_1 \, V &+ a_2 \, V^2 &+ \ldots\text{ ,}\label{eqn_tsl_bias}
		\\
		&T_\mathrm{OOP}{}(V) \, &= \, b_1 \, V &+ b_2 \, V^2 &+ \ldots\text{ .}\label{eqn_tfl_bias}
	\end{alignat}
\end{subequations}

The fixed parameters used in all simulations are $\alpha = 0.01$, an elliptical free layer with a minor axis of 70~nm and a major axis of 170~nm, $\Delta t = 5$~ps, total integration time $t_\mathrm{end} = 1$~ms, a small initial angle between $\vect{m}$ and $\vect{p}$ of $1$~mrad, and temperature $T = 300$~K. The effect of Joule heating can be neglected in our samples.\citep{georges_2009_origin,chanthbouala_2011_vertical-current-induced} The LLG equation~(\ref{eqn_llg_stt}) is then solved using the classical fourth-order Runge--Kutta method. For each numerical integration, the final state reported in the switching phase diagrams corresponds to time-averaged values at the end of each simulation. In this way, oscillatory states result in intermediate, non-saturated values as it is the case in dc measurements.

\begin{figure}[bt]
	\includegraphics[width=0.98\columnwidth]{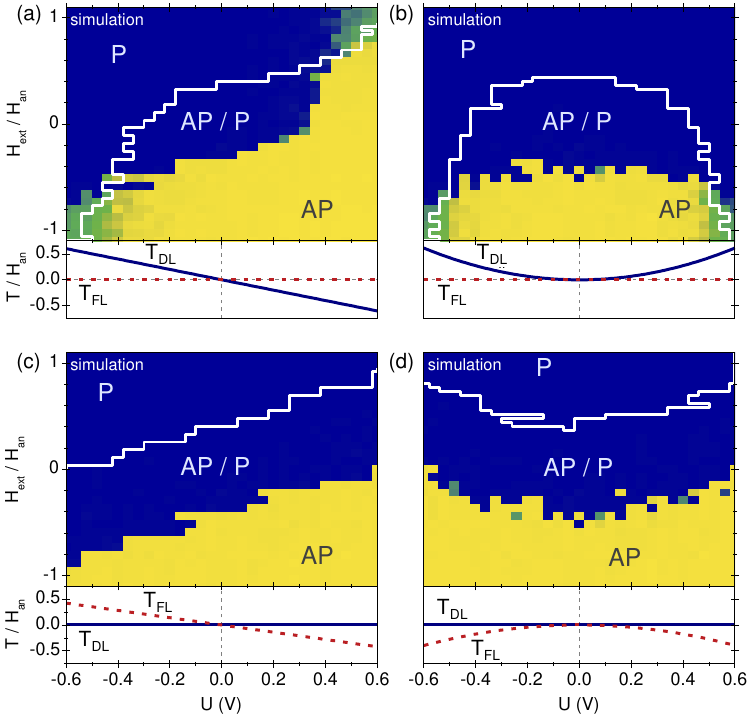}%
	\caption{\label{fig:figure1} (Color online) Simulated switching phase diagrams and corresponding spin-torques as a function of voltage. Colors indicate magnetization state where dark (blue) regions corresponds to parallel (P) and bright (yellow) regions to antiparallel (AP) alignment of the fixed and free layer magnetizations. The results shown here are obtained for increasing field values; the white line marks the switching border in the case of decreasing field. Each panel shows the effect of one bias-dependence of spin-torque at a time: (a)~Linear damping-like torque. (b)~Quadratic damping-like torque. (c)~Linear field-like torque. (d)~Quadratic field-like torque.}
\end{figure}

As shown in \reffig{fig:figure1}, different bias-dependences of damping-like and field-like torques result in unique features in the switching phase diagrams. First, the influence of the two types of spin-torques is fundamentally distinct. As the applied bias increases, the damping-like component eventually closes the bistable area in which parallel (P) and antiparallel (AP) configurations of the fixed and free-layer magnetizations coexist. In contrast, the field-like torque directly adds up to the external magnetic field and results in a shift of the hysteresis curve as a function of applied bias. Therefore, the bias-dependence of the field-like torque directly shows up in the switching phase diagram [\refsfig{fig:figure1}{c and d}]. At higher voltages, the damping-like torque leads to sustained magnetization oscillations, resulting in a reduced averaged magnetization amplitudes [\refsfig{fig:figure1}{a and b}]. This allows for an unambiguous determination of the respective torque amplitudes for which we adapt the coefficients of the torques in \refeq{eqn_stt_bias} until the simulated phase diagrams match the experimental ones. In order to estimate the uncertainty of our results on the bias dependent spin-torques, we vary each coefficient [$a_i$ resp. $b_i$ in \refeq{eqn_stt_bias}] to the point where the agreement between simulation and measurement is palpably lost. The errors on the final results that are given in \refsfig{fig:figure2}{m--o} are calculated from the combination yielding the highest error for each bias. As we are only interested in the bias-dependences of the torques and not in their absolute amplitudes, the assumption of a fully spin-polarized current does not represent any restriction.

\section{Experiments}

\begin{figure*}[bt]
	\includegraphics{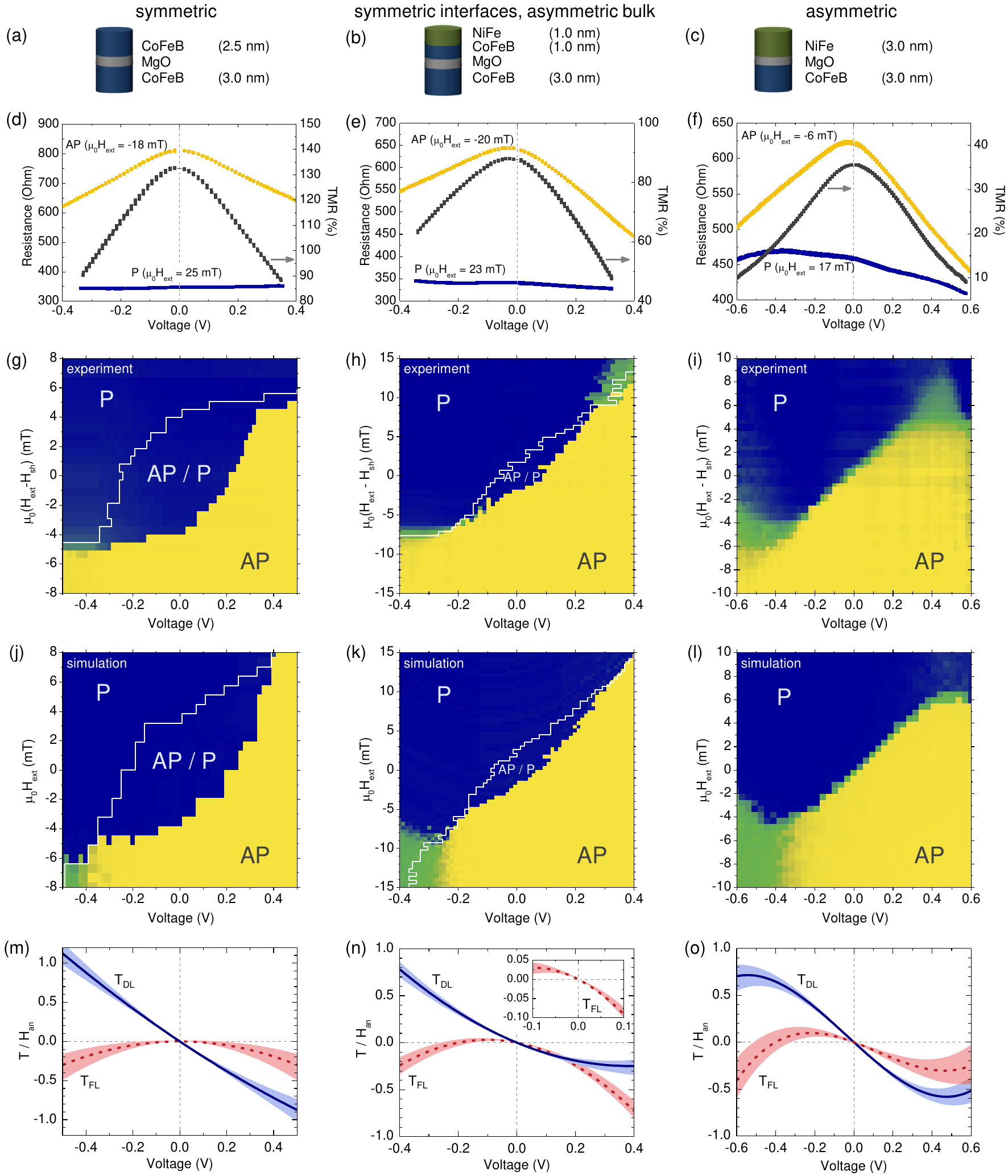}%
	\caption{\label{fig:figure2} (Color online) (a--c) Simplified layer stacks of the samples used in this study. (d--f) Voltage dependent resistances of the saturated parallel (P) and antiparallel (AP) states and corresponding TMR values. Experimental (g--i) and simulated (j--l) switching phase diagrams. The white lines indicate the border for AP to P state switching for increasing applied fields. (m--o) Resulting bias-dependent spin-torques as used in the simulations.}
\end{figure*}

The samples have a common layer stack of CoFe(2.5)/Ru(0.85)/Co$_{60}$Fe$_{20}$B$_{20}$(3)/MgO(1.1)/free layer (thicknesses in nanometers). In order to elucidate the influence of asymmetry in electrode composition, samples with three different compositions of the top electrode, that forms the free layer, have been fabricated [see \refsfig{fig:figure2}{a--c}]. The films were deposited by sputtering. Then the junctions were patterned into an elliptical shape (170~nm $\times$ 70~nm) by ion milling in order to introduce an easy magnetization axis thanks to shape anisotropy. A more detailed description of the fabrication process is given in Ref.~\onlinecite{nagamine_2006_ultralow}.

The experiments consist of dc resistance measurements while sweeping an external magnetic field parallel to the easy axis at a certain fixed applied voltage bias. Under positive voltage, the electrons flow from the free to the fixed layer and a positive magnetic field favors parallel alignment of fixed and free layer magnetizations. In order to remove the strong voltage-dependence of the resistance, we use normalized values in the diagrams: $R_\mathrm{rel}(V,H_\text{ext}) = [R(V,H_\text{ext}) - R_\mathrm{P}(V)] / [R_\mathrm{AP}(V) - R_\mathrm{P}(V)]$, where $R_\mathrm{P}(V)$ and $R_\mathrm{AP}(V)$ are fully saturated P and AP states, respectively, stabilized by high external magnetic fields [see \refsfig{fig:figure2}{d--f}]. As the measurements are taken on dc time-scales, the free layer magnetization is expected to be in equilibrium state. The results for a range of bias values are represented in switching phase diagrams [see \refsfig{fig:figure2}{g--i}]. In these diagrams, intermediate resistance values between the minimum and maximum values of the static P and AP states correspond to dynamic oscillatory states. A small field shift $H_\mathrm{sh}$ in the low-bias hysteresis curves has been subtracted from the presented data. It most certainly results from the dipolar coupling to the reference layer due to an incomplete screening. This subtraction does not influence our bias-dependent results, but implies that we ignore any zero-bias interlayer exchange coupling.\citep{slonczewski_1989_conductance,bruno_1999_theory}

\section{Results and discussions}

\subsection{Symmetric junction}

We first performed measurements on samples with symmetric electrode interfaces, both composed of the same material (Co$_{60}$Fe$_{20}$B$_{20}$) giving rise to TMR values of about 130\% at low bias [see \refsfig{fig:figure2}{d}]. The symmetry of the junction is also translated into highly symmetric bias-dependences of the saturated P and AP states resistances [see \refsfig{fig:figure2}{d}]. The experimentally obtained switching phase diagram [\refsfig{fig:figure2}{g}] shows a fully closed hysteretic zone at the center and is mainly symmetric. This can also be observed in the bias-dependent resistance for the satured states in \refsfig{fig:figure2}{d}. These measurements could be reproduced by our simulation to a very good agreement [\refsfig{fig:figure2}{j}, using a saturation magnetization of $M_\text{s} = 1.38$~T, Ref.~\onlinecite{kubota_2006_dependence}]. 

The resulting spin-torques [\refsfig{fig:figure2}{m}] show the expected bias-dependences of a symmetric magnetic tunnel junction:\citep{heiliger_2008_ab} the damping-like torque is a strongly linear function of voltage with a small quadratic contribution. In contrast, the field-like torque is purely quadratic and therefore a symmetric function of the applied bias. These results agree very well with other results obtained with symmetric magnetic tunnel junctions by both experimental\citep{kubota_2007_quantitative,sankey_2007_measurement,wang_2009_bias,jung_2010_bias} and theoretical\citep{theodonis_2006_anomalous,xiao_2008_spin-transfer,manchon_2008_description,wilczynski_2008_free-electron,kalitsov_2009_spin-transfer,franz_2013_influence} techniques. They also validate our method for extracting the spin-torque bias-dependences by matching experimental and simulated phase-diagrams.

\subsection{Symmetric interfaces, asymmetric bulk}

The second sample with a free layer of Co$_{60}$Fe$_{20}$B$_{20}$ (1~nm)/Ni$_{81}$Fe$_{19}$ (1~nm) [see \refsfig{fig:figure2}{b}] possesses symmetric interfaces at the MgO barrier but different bulk materials in the electrodes resulting in a slight asymmetry in the bias-dependence of the TMR [\refsfig{fig:figure2}{e}]. The experimental switching phase diagram resembles that obtained on the fully symmetric sample [\refsfig{fig:figure2}{h}]. However, the hysteretic zone shows a stronger, linear shift with voltage which indicates a non-vanishing linear component in the bias-dependence of the field-like torque. We were able to reproduce the experimental result by simulation with only a small deviation at high negative voltages [\refsfig{fig:figure2}{k}, using the mean of the saturation magnetizations of CoFeB\citep{kubota_2006_dependence} and NiFe\citep{finocchio_2007_micromagnetic} $M_\text{s} = 1.11$~T]. The strong oscillations observed in the simulations are most probably suppressed by micromagnetic effects in our samples. The best agreement to measurements is achieved under the assumption of an increased quadratic damping-like torque in combination with a clearly linear bias-dependence of the field-like torque at small voltages [\refsfig{fig:figure2}{n} and inset therein]. At higher bias it finally recovers its quadratic form as observed in the symmetric junction.

\subsection{Asymmetric junction}

In the case of a free layer of Ni$_{81}$Fe$_{19}$ (3~nm) [see \refsfig{fig:figure2}{c}], the bias-dependent P and AP state resistances reflect the strong asymmetry in the junction's structure [see \refsfig{fig:figure2}{f}]. The switching phase diagram features strong differences from those obtained for symmetric interfaces [\refsfig{fig:figure2}{i}]. Additionally, at room temperature, the sample is superparamagnetic, not displaying any hysteresis, which can be ascribed at least partially to its lower saturation magnetization of $M_\text{s} =0.81$~T.\citep{finocchio_2007_micromagnetic}

In order to reproduce the phase diagram [\refsfig{fig:figure2}{l}], it is necessary to include a third order voltage term into the description of the spin-torques' bias-dependences. The switching border is mainly influenced by the field-like torque which is a nearly linear function of voltage at small bias. A similar dependence has been measured in other (slightly) asymmetric magnetic tunnel junctions\citep{oh_2009_bias-voltage,chanthbouala_2011_vertical-current-induced,matsumoto_2011_spin-torque} Theoretical calculations also yield this linear component.\citep{tang_2009_controlling,manchon_2010_signatures,tang_2010_influence} At higher voltages, however, the evolution deviates from the previous results. The bias-dependence is twisted and reigned by a third order term. The damping-like component also exhibits a mainly linear bias-dependence at small voltage amplitude as it is measured in symmetric junctions. However the apparition and subsequent suppression of oscillations, observable through the intermediate relative resistance values at voltages amplitudes of $\pm$0.4~V, can only result from a reduction of the damping-like torque amplitude at high absolute voltage values. These assumptions are in good agreement with the results of the bias-dependent torques [\refsfig{fig:figure2}{o}].

Higher orders voltage terms in the bias-dependence have also resulted from the calculations by \citet{tang_2010_influence}, modeling the asymmetry of the junction by a parameter describing a shift in the spin-dependent on-site energies of the two ferromagnetic electrodes. \citet{kalitsov_2013_spin} have shown similar results by introducing different electrode interfaces. However, these theoretical results are linked to negative TMR values appearing in the low bias region which we did not observe in our sample [see \refsfig{fig:figure2}{f}]. In summary, although many models are in a good partial agreement with our experimental results, a full understanding and theoretical description of the bias-dependence of spin-torques in asymmetric junctions is still missing.

\section{Conclusions}

We have developed a new method to extract the bias-dependences of the damping-like and field-like spin-torques in MgO-based magnetic tunnel junctions without any assumption on their variations. A symmetric junction with identical electrodes of CoFeB exhibits a linear damping-like and a quadratic field-like torque, as expected from theory. By introducing an asymmetry in the bulk of an CoFeB(1~nm)/NiFe(1~nm) free-layer, both torques are modified: the damping-like component gains a small quadratic dependence and the field-like torque becomes linear at small bias. For the case of a completely asymmetric junction with a NiFe free layer we found that the bias-dependences are twisted. Both spin-torques acquire higher order terms and the amplitude of the field-like torque becomes similar to that of the damping-like torque. Although this has been partially predicted by theoretical calculations, more experimental results are required to fully understand the parameters that influence spin-torques in magnetic tunnel junctions. The technique we developed in this manuscript allows an easy-to-implement and fast way to extract the bias-dependence of torques in any kind of sample. This will eventually allow the systematic design of magnetic tunnel junction structures for specific applications.


\begin{acknowledgments}
	The authors would like to thank Aur\'elien Manchon and Mairbek Chshiev for useful discussions, and Canon ANELVA for preparation of the magnetic films. Financial support from the European Research Council (Starting Independent Researcher Grant No.\ ERC 2010 Stg 259068) is acknowledged.
\end{acknowledgments}

\bibliography{bibliography}

\end{document}